%% file: main.tex
\documentclass[runningheads]{llncs}

\usepackage{cite}
\usepackage[utf8]{inputenc}
\usepackage{lineno}
\usepackage{calc}
\usepackage{etoolbox}
\usepackage{tikz}
\usepackage{amsmath}
\usepackage{amssymb}
\usepackage{amsfonts}
\usepackage{mathtools}
\usepackage{wrapfig}
\usepackage{verbatim}
\usepackage{hyperref}
\usepackage[ruled, linesnumbered]{algorithm2e}

\usetikzlibrary{shapes.geometric, arrows, calc, automata}

\newcommand{\false}{\textit{false}}

\newcommand{\tab}[2]{\setlength{\hspace*}{0.4cm * #2}#1\setlength{\hspace*}{0.4cm - \widthof{#1}}}

\newcommand{\pt}{\textit{pt} = \langle a, q, \textit{next}, d, v, v_{\textit{max}} \rangle}

\SetKwBlock{DoEither}{do either}{end}
\SetKwBlock{OrDo}{or do}{end}
\SetKwIF{Try}{On}{Finally}{try}{}{on}{finally}{end}
\DontPrintSemicolon

\title{Efficient dynamic model based testing
\thanks{This publication is part of the PVSR project (with project number 17933) of the MasCot research programme which is financed by the Dutch Research Council (NWO).}\\ 
\small{using greedy test case selection}
}

\author{P.H.M.\ van Spaendonck \orcidID{0000-0002-9536-1524}}

\institute{Department of Mathematics and Computer Science,\\
    Eindhoven University of Technology\\
    \texttt{P.H.M.v.Spaendonck@tue.nl}}

\titlerunning{Efficient dynamic model based testing}

\begin{document}

\maketitle
\begin{abstract}
Model-based testing (MBT) provides an automated approach for finding discrepancies between software models and their implementation. 
If we want to incorporate MBT into the fast and iterative software development process that is Continuous Integration Continuous Deployment, then MBT must be able to test the entire model in as little time as possible.

However, current academic MBT tools either traverse models at random, which we show to be ineffective for this purpose, or use precalculated optimal paths which can not be efficiently calculated for large industrial models.
We provide a new traversal strategy that provides an improvement in error-detection rate comparable to using precalculated paths.
We show that the new strategy is able to be applied efficiently to large models.
The benchmarks are performed on a mix of real-world and pseudo-randomly generated models.
We observe no significant difference between these two types of models.
\end{abstract}

\input{sec1-introduction/sec1-main.tex}
\input{sec2-theory/sec2-main.tex}

\input{sec3-mbt-algorithm/sec3-main.tex}
\input{sec4-greedy-strategy/sec4-main.tex}
\input{sec6-benchmarks/sec6-main.tex}
\input{sec7-conclusion/sec7-main.tex}

\bibliographystyle{splncs04}
\bibliography{bibliography}

\end{document}

%% file: sec1-introduction/sec1-main.tex
\section{Introduction}
    Testing has become a core tenet of modern-day software engineering.
    It has been repeatedly shown that having thoroughly tested software leads to higher quality software with significantly lower maintenance and development costs \cite{TDD, beck2003test}.
    This has led to test-driven engineering, in which tests for a component or feature are written before its actual implementation commences, and the implementation is deemed to be correct only once all tests succeed.
    
    Similarly, we see a rise in the usage of model-driven engineering, in which software components are first described on a higher level of abstraction as a state machine/model.
    Model-based techniques can then be used to deliver correct and verified software.
    For example, in \cite{BJN07}, UMLsec models are used to verify that the security requirements of a single-sign-in software application are correctly enforced.
    
    Model-Based Testing (MBT), such as the \textbf{ioco}-based approach originally outlined in \cite{IOCO} by Tretmans, sits at the intersection of these two engineering approaches and allows us to use an abstract model to automatically test its implementation.
    The ability to automatically generate tests makes MBT very useful for software development by, in theory, removing the need to spend time manually writing tests.
    In practice, however, the exhaustive approach of MBT can easily lead to testing the same behavior multiple times.
    This contrasts with manually written test suites which are often designed to have as little overlap in tested behavior as possible.
    Reducing the overlap in tested behavior when using MBT should thus be considered a critical step for applying it for the testing of large complex systems.

    The focus of our research is reducing overlap in tested behavior when applying the \textbf{ioco}-based MBT technique. 
    For this technique, we confine ourselves to discrete datatypes and we formalize systems as directed graphs in which edges can be labeled with inputs or outputs that use these datatypes.
    As such, a test run through the model should have as little overlap with itself as possible.
    Visiting the same part of the model multiple times means that the behavior corresponding to that part of the model would also be tested multiple times.
    
    In \cite{PetraTretmans2019}, van den Bos and Tretmans provide a possible solution to reducing this overlap by automatically calculating test paths that cover all symbolic transitions of a given model.
    The MBT tool then uses an SMT solver to calculate concrete values for the queries while traversing the pregenerated path.
    Their approach shows a significant speedup over the classic random test-case selection approach when used to detect non-conformances in several mutants of the Bounded Retransmission Protocol \cite{BoundedRetransmissionProtocol}.

    However, calculating such a global optimal path is \textit{NP}-hard and thus becomes computationally too expensive when applied to large industrial systems.
    Another issue, which van den Bos and Tretmans also acknowledge, is that using the same test path every time might cause the test to miss bugs that do not present themselves along the given path \cite{PetraTretmans2019}.
    Likely because of these reasons, academic MBT tools, such as TorXakis \cite{Tretmans2019}, opt to traverse the model at random instead.
    
    We investigate whether a simple and straightforward greedy test-case selection strategy can be used to attain a similar speedup to the one attained through the usage of pre-generated paths, whilst also being able to be applied to large models.
    For the initial comparison, we use our greedy strategy to run the same experimental setup used for the global optimal strategy in \cite{PetraTretmans2019}, i.e.\ mutation detection on the Bounded Retransmission Protocol \cite{BoundedRetransmissionProtocol}.
    Our greedy strategy performs $8.2$ times better than the random strategy according to the arithmetic mean.

    Second, we test the scalability of the greedy solution by applying it to large real-world models, as well as randomly generated statespaces of varying sizes.
    These random statespaces are approximations of complex systems running under run-to-completion semantics.
    The greedy strategy provides a noticeable speedup over the random test-case selection strategy when aiming to cover at least $70\%$ of the states of the given models.
    We also observe that the generated models provide similar results to those of the real-life use case.

    In Section \ref{sec:related_work}, we briefly discuss related work on model-based testing techniques and optimizations.
    In Section \ref{sec:theory}, we give the underlying formalization of the theory of \textbf{ioco}-based MBT.
    In section \ref{sec:algorithm}, we discuss the algorithm used for \textbf{ioco}-based MBT.
    In Section \ref{sec:optimization}, we discuss our new optimizations.
    In Section \ref{sec:results}, we show and discuss our benchmarks.
    We give our final conclusion and discuss other areas of research that might be important for the usage of MBT within the software development cycle in Section \ref{sec:fin}.

\section{Related work} \label{sec:related_work}
    When formalizing systems as abstract models, we often prefer abstracting away over data and formalizing them as discrete datatypes, as this significantly reduces the complexity of the system.
    However, when formalizing low-level cyber-physical components, we see that the exact value of data, e.g.\ the angle of an air-intake vent, can not be abstracted away over, as they are central to the correctness of such components.
    In such cases, systems can be formalized as continuous or hybrid automata \cite{Henzinger2000}.
    We highlight some of the state-of-the-art test optimization techniques used for systems with continuous data.

    For systems with continuous data, MBT is done through search-based testing in which meta-heuristics, such as genetic algorithms or simulated annealing, are used to automatically generate test data.
    Reducing overlap in tested behavior can, in these instances, be reformulated as an optimization problem.
    As an example, in \cite{SearchOptimization}, meta-heuristics are used to find optimal inputs that maximize the distance between the expected optimal input, thus increasing the likelihood of a fault occurring, and between previous inputs, decreasing the possible overlap with previously tested behavior.
    In \cite{MiL-spacereduction}, a combination of dimensionality reduction and surrogate modeling techniques based on supervised learning is used to scale up similar techniques to be able to be applied to the incredibly large search spaces encountered within the cyber-physical domain of the automotive industry.
    
    The systems that we describe in Section \ref{sec:theory} and use throughout the paper, are models with discrete datatypes, often with no direct interrelation.
    As such, there is no continuous search space to optimize over and the test case selection comes down to optimizing the traversal of the graph that is the specification model.
    However, these two techniques are not exclusive, and can very well be combined, such as is shown in \cite{funcsearch}, in which search-based techniques are used to pick viable candidates for the data parameters of functions.

    Last, we discuss work on reducing overlap in tested behavior when using online, i.e.\ a finite test suite is generated a priori, model-based conformance testing techniques.
    In \cite{Cartaxo11}, Cartaxo et al.\ provide a solution to reducing overlap in tested behavior by measuring the similarity of different tests in the generated test suite and picking a subset of tests such that the similarity measure between the tests in the subset is as low as possible.
    By doing so, they are able to reduce the number of run tests with $80\%$ while still maintaining a similar fault detection rate to that of a full run test suite.
    In \cite{AICHERNIG2015383}, Aichernig et al.\ use model-based mutation testing in which mutant models i.e.\ models containing faulty behavior, are generated from the correct specification model, which, in turn, are used to generate tests aimed at detecting the faults contained in the mutant models.
    Aichernig et al.\ use a bounded equivalence checker to find and remove identical mutants that would lead to tests being generated with an overlapping tested behavior.
    Offline MBT techniques, such as the one used by us, are generally more suited for testing behavior that only occurs after long specific sequences of inputs and outputs.

%% file: sec2-theory/sec2-main.tex
\section{Theory of model based testing} \label{sec:theory}
    In this section, we give the theory needed to understand how the behavior of a given system is formalized, and how these formal models are used during dynamic MBT.
    The MBT theory presented here is the same as the original theory described in \cite{IOCO} with only some notational differences. 
    
    The expected behavior of a system is expressed as an input-output labeled transition system (IOLTS), a variation of labeled transition systems in which the action labels are split into disjoint input- and output-action sets. 
    We define an IOLTS as follows:
    
    \begin{definition}
        An IOLTS $L$ is defined as the 5-tuple $L = \langle Q,q_0,A_I,A_O,\to \rangle$, where:
        \begin{itemize}
            \item $Q$ is the finite set of states,
            \item $q_0 \in Q$ is the initial state,
            \item $A_I$ and $A_O$ are the sets of input- and output-actions, respectively, such that $A_I \cap A_O = \emptyset$ and $\tau, \delta \not\in A_I \cup A_O$.
            \item $\to \; \subseteq Q \times (A_I \cup A_O \cup \{\tau, \delta\}) \times Q$ is the transition relation, such that $\langle q, \delta, q'\rangle \in \to $ iff
            $q = q'$ and the state $q$ has no outgoing transitions labeled with an output action $a_{\textit{out}} \in A_O$ or $\tau$-action.
        \end{itemize}
    \end{definition}
    Given states $p,q$, and some action label $a$, we use the shorthand notation $p \overset{a}{\to} q$ instead of $\langle p,a,q \rangle \in \to$.
    
    In the set of actions $A_I \cup A_O$, each element corresponds to a function in the implementation.
    We make a distinction between input actions $A_I$, functions that we call as inputs to the system, and output actions $A_O$, functions of other components that can either trigger as a response to inputs given to the system, or spontaneously, e.g.\ by some internal timer timing out. 
    
    We further extend our set of actions with the special actions $\tau$ and $\delta$.
    The $\tau$-action represents an externally non-observable action, e.g.\ some internal calculation that occurs as a result of calling a function.
    The $\delta$-action, referred to as quiescence, represents the system remaining idle.
    A transition with a $\delta$-action occurs exactly in all states in which it is not possible to take an output transition nor an internal action and must be a self-loop, i.e.\ the begin- and endpoint must be the same.

    For readability, we write $A_\delta$ as shorthand for $A_I \cup A_O \cup \{\delta\}$ and we write $A$ as shorthand for $A_I \cup A_O$.

    During MBT, a sequence of actions referred to as a \textit{suspension trace}, is used to keep track of the actions performed thus far.
    We define the set of suspension traces using Definitions \ref{def:path_to}, and \ref{def:traces}.
    
    \begin{definition} \label{def:path_to}
        Let $L = \langle Q, q_0, A_I, A_O, \rightarrow\rangle$ be an IOLTS. We define the observable path relation $\Rightarrow\, \subseteq Q \times A_{\delta}^* \times Q$ as the smallest relation satisfying:
        \begin{itemize}
            \item $q \overset{\epsilon}{\Rightarrow}q$ given any state $q\in Q$, where $\epsilon$ refers to the empty sequence,
            \item given states $q,q',q'' \in Q$ and some word $w\in A_{\delta}^*$,\\ if $q\overset{\tau}{\to}q'$ and $q' \overset{w}{\Rightarrow}q''$ then $q \overset{w}{\Rightarrow}q''$, and
            \item given states $q,q',q'' \in Q$, an observable action $a\in A_{\delta}$, and a word $w\in A_{\delta}^*$,
            if $q\overset{a}{\to}q'$ and $q' \overset{w}{\Rightarrow} q''$ then $q\overset{aw}{\Rightarrow}q''$.
        \end{itemize}
    \end{definition}
    
    \begin{definition} \label{def:traces}
        Let $L = \langle Q, q_0, A_I, A_O, \rightarrow \rangle$ be some IOLTS. Given a state $q\in Q$, the set of suspension traces $\textit{straces}(q) \subseteq A_\delta^*$ is defined as follows:
        \[\textit{straces}(q) = \{ w \in A_\delta^* \;|\; \exists_{q'{\in} Q} [ q \overset{w}{\Rightarrow} q']\} \text{.}\]
    \end{definition}
    
    Model Based Testing tests whether the behavior of the implementation conforms to the behavior of the specification, i.e.\ the implementation never gives an output that the specification does not allow.
    This conformance is described using the \textbf{i}nput-\textbf{o}utput-\textbf{co}nformance relation \textbf{ioco} on IOLTSs, originally defined in \cite{IOCO}, and is given in Definition \ref{def:ioco}.

    \begin{definition} \label{def:after_out}
    Let $L = \langle Q, q_0, A_I, A_O, \to \rangle$ be an IOLTS. We define the mappings $\textit{out}_L : \mathcal{P}(Q) \to \mathcal{P}(A_O)$ and $L~\textbf{after} : A_\delta^* \to \mathcal{P}(Q)$ as follows:
    \begin{itemize}
            \item given some set of states $\textit{qs} \subseteq Q$, we have
            \[\textit{out}_L(qs) = \{a_{\textit{out}} \in A_O | \exists_{q{\in}\textit{qs},q'{\in}Q}[ q \xrightarrow{a_{\textit{out}}}q']\}  \text{, and}\]
            \item given some suspension trace $\sigma \in \textit{straces}(q_0)$, we have
            \[ L~\textbf{after}~\sigma = \{ q \in Q | q_0 \overset{\sigma}{\Rightarrow} q\} \text{.}\]
        \end{itemize}
    \end{definition}
    
    \begin{definition} \label{def:ioco}
        Given an \textit{IOLTS} $L_{\textit{impl}}$ of the implementation and an \textit{IOLTS} $L_{\textit{spec}} = \langle Q, q_0, A_I, A_O, \to \rangle$ of the specification, the conformance $L_{\textit{impl}}~\textbf{ioco}~L_{\textit{spec}}$ holds iff given any trace $\sigma \in \textit{traces}(q_0)$, we have
            \[ \textit{out}_{L_{\textit{impl}}}(L_{impl}~\textbf{after}~\sigma ) \subseteq \textit{out}_{L_{\textit{spec}}}(L_{\textit{spec}}~\textbf{after}~\sigma) \text{.}\]
    \end{definition}

%% file: sec3-mbt-algorithm/sec3-main.tex
\section{The MBT algorithm} \label{sec:algorithm}
\newcommand{\nexta}{\texttt{next}(L, qs)}
    We now discuss the MBT algorithm from \cite{IOCO}, for which we give a pseudocode description in Algorithm \ref{tbl:MBT_algorithm}.
    The MBT algorithm uses the set $qs$, which initially contains the initial state $q_0$ and all states reachable from there using only $\tau$-actions, to keep track of which states it could be in at the start of each iteration.
    We define $\nexta$ to be the set of possible actions that can be taken from at least one state in $qs$, i.e.\  $\nexta = \{a \in A_{\delta} | \exists_{q{\in}\textit{qs},q'{\in}Q}[ q \overset{a}{\to}q']\}$.

    \begin{algorithm}
    \caption{MBT algorithm}
    \label{tbl:MBT_algorithm}

    \KwData{A specification IOLTS $L=\langle Q, q_0, A_I, A_O, \rightarrow \rangle$ and a timeout time $t$.}
    
    $qs \gets \{ q \in Q | q_0\overset{\epsilon}{\Rightarrow} q\}$\;   
    \textbf{while}~$\nexta \neq \emptyset$ \textbf{do :}\;
    \tab{}{1}$\textbf{do either}:$\;
    \tab{}{2}$\textbf{if}~\nexta \cap A_I \not= \emptyset~\textbf{then}:$ \;
    \tab{}{3}\textbf{select} $a_{\textit{in}}$ \textbf{from} $\nexta \cap A_I$ \;
    \tab{}{3}\texttt{send}$(a_{\textit{in}})$ \;
    \tab{}{3}$qs \gets \{ q \in Q | \exists q' \in qs : q'\xRightarrow{a_{\textit{in}}}q\}$ \;
    \tab{}{1}$\textbf{or}:$ \;
    \tab{}{2}\textbf{try} $a_{\textit{out}} \gets \texttt{rcv}(t)$ \;
    \tab{}{3}\textbf{if} $a_{\textit{out}} \in \nexta$ \textbf{then} : \;
    \tab{}{4}$q \gets \{ q \in Q | \exists q' \in qs : q' \xRightarrow{a_{\textit{out}}} q\}$ \;
    \tab{}{3}\textbf{else} : \;
    \tab{}{4}\textbf{return} $\false$ \;
    \tab{}{2}\textbf{on} \textit{timeout} : \;
    \tab{}{3}\textbf{if} $\delta \in \nexta$ \textbf{then} : \;
    \tab{}{4}$qs \gets \{ q \in qs | q \overset{\delta}{\rightarrow}q\}$ \;
    \tab{}{3}\textbf{else} : \;
    \tab{}{4}\textbf{return} $\false$ \;
    \end{algorithm}

    The function $\texttt{send}(a)$ on line 6 causes the implementation to execute the function corresponding to the input action $a$. 
    Whenever the implementation calls a function corresponding to some output action $a_{\textit{out}}$, the label $a_{\textit{out}}$ is sent back to the MBT algorithm and added to a response queue.
    Messages in the queue are read first-in-first-out using the $\texttt{rcv}(t)$ function on line 9, removing the read message in the process. 
    If $\texttt{rcv}(t)$ is called and the queue is empty, it will wait till a new message is received.
    If no message is received within the timeout time $t$, the function throws a \textit{timeout}.
    
    The algorithm repeatedly does one of two things:
    \begin{itemize}
        \item Some input action $a_{\textit{in}}$ that is possible from any state in $qs$ is picked and the implementation is requested to execute the corresponding function using $\texttt{send}(a_{\textit{in}})$.
        The tool then updates the set of possible states $qs$ accordingly and continues.
        \item The MBT tool listens for a possible output action $a_{\textit{out}}$ using $\texttt{rcv}(t)$.
        If an action is received, we verify whether it is an allowed action.
        If the action is allowed, the set of states $qs$ is updated accordingly and we continue.
        If $a_{\textit{out}}$ is not a possible action from any state in $qs$, then we can conclude that the implementation is not \textbf{ioco} with the specification model, and the test has failed. 
        If no response is received within the timeout time $t$, we assume the implementation to be idle/quiescent.
        The algorithm reduces the set $qs$ to only the quiescent states in $qs$, i.e.\ states with a $\delta$-loop.
        If none of the states in $qs$ are quiescent, then the implementation is also not \textbf{ioco} with the specification model, and the test has failed.
    \end{itemize}
    
    Note that the MBT algorithm does not terminate if no behavioral differences are ever detected. 
    The tester must decide when to terminate.
    The termination criterion that is used throughout the paper is state coverage, i.e.\ the percentage of states that have been reached/tested throughout the run.

%% file: sec4-greedy-strategy/sec4-main.tex
\section{The greedy test case selection strategy} \label{sec:optimization}
    To reduce the amount of overlap in tested behavior, we want to avoid querying inputs leading to already tested states.
    However, calculating a global optimal path, i.e.\ finding the shortest path visiting each state at least once, is too computationally expensive to be done on complex systems which have a large number of states.
    Instead, our proposed strategy intends to approximate the global optimal path by picking locally optimal solutions.
    This is done by, calculating all possible paths originating from our current set of states $\textit{qs}$, of a given length $n$, and then picking the input action corresponding to the path with the least states that have already been visited.
    If more than one such action is available, a random contender is picked.

    We make use of two optimizations to reduce the amount of work required to find these local-optimal paths.
    To highlight the need for these optimizations, let us consider an IOLTS in which each state has, on average, $\lambda$ outgoing transitions.
    Simply calculating all possible paths of length $n$ and picking the most optimal one, would still be in the order of $O(|\textit{qs}|\cdot \lambda^n)$.

    For the first optimization, we note that once an optimal path of length $n$ has been found, and a singular action has been performed, it is unnecessary to reconsider all possible paths of length $n$, since we already have a near-optimal path of length $n-1$.
    Instead, we only consider the extensions of our leftover path.
    In Definition \ref{def:path-tree} we outline the path-tree data structure that we use to keep track of the previously performed calculations.

    \begin{definition} \label{def:path-tree}
    Given an IOLTS $L = \langle Q, q_0, A_I, A_O, \to \rangle$, a path-tree is a 6-tuple $\pt$, where
    \begin{itemize}
        \item we have projection functions $\FuncSty{a},\FuncSty{q},\FuncSty{next},\FuncSty{d},\FuncSty{v}$ and $\FuncSty{vmax}$ defined on \textit{pt} such that $\FuncSty{a}(\textit{pt}) = a$, $\FuncSty{q}(\textit{pt}) = q$, $\FuncSty{next}(\textit{pt}) = \textit{next}$, $\FuncSty{d}(\textit{pt}) = d$, $\FuncSty{v}(\textit{pt}) = v$, and $\FuncSty{vmax}(\textit{pt}) = v_{\textit{max}}$,
        \item $a$ is an action label leading to state $q$, i.e.\ $\exists_{ q'\in Q } [q' \xrightarrow{a} q]$,
        \item \textit{next} is a set of path-trees such that $\forall_{ \textit{pt}' {\in} \textit{next} } [q \xrightarrow{\FuncSty{a}(\textit{pt}')} q']$,
        \item $d$ is the depth of the tree, i.e.\ $1$ plus the highest depth among the path-trees in \textit{next},
        \item $v$, i.e.\ the value of a path-tree, equals the maximum amount of unvisited states that can be reached through a sequence of $d$ transitions, starting with the transition $q' \xrightarrow{a}q$, and
        \item $v_{\textit{max}}$ is used to store the maximum value among the path-trees in $\textit{next}$.
    \end{itemize}
    \end{definition}

    \begin{algorithm}
        \KwData{The IOLTS $L= \langle Q, q_0, A_I, A_O, \rightarrow \rangle$, a path-tree $\pt$, and a target depth of $n \geq d$.}
        \tab{}{0} \textbf{if} $d = n$ :\;
        \tab{}{1} \textbf{return;}\;
        \tab{}{0} \textbf{else if} $\textit{next} = \emptyset$:\;
        \tab{}{1} \textbf{for} $a{:}A, q'{:}Q$ \textbf{s.t.} $q\overset{a}{\to} q'$ \textbf{do:}\;
        \tab{}{2} \textbf{add} $\langle a, q', \emptyset, 1, \textit{covered}(q'), 0 \rangle$ \textbf{to} $\textit{next}$\;
        \tab{}{0} \textbf{for} $\textit{pt}' \in \textit{next}$ \textbf{do:}\;
        \tab{}{1} \textbf{if} $\FuncSty{v}(\textit{pt}') + ((n - 1) - \FuncSty{d}(\textit{pt}')) \geq v_\textit{max}$ \textbf{:}\;
        \tab{}{2} \texttt{grow}$(L, \textit{pt}', n-1)$\;
        \tab{}{2} $v_{\textit{max}} \gets \texttt{max}(v_{\textit{max}}, \FuncSty{v}(\textit{pt}'))$\;
        \tab{}{0} $v \gets v_{\textit{max}} + \textit{covered}(q)$ \;
        \tab{}{0} $d \gets n$ \;
    \caption{The $\texttt{grow}(L, \textit{pt}, n)$ algorithm for path-trees}
    \label{tbl:path-tree}
    \end{algorithm}

    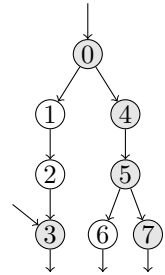
\begin{wrapfigure}{r}{0.2\textwidth}
        \centering
        \begin{tikzpicture}
        \tikzstyle{tested}=[draw, circle, fill=gray!20, inner sep=1pt]
        \tikzstyle{untested}=[draw, circle, inner sep=1pt]

        \node (null) at (0,.8) {};
        
        \node[tested] (0) at (0,0) {$0$};
        \draw[->] (null) -- (0);
        
        \node[untested] (1) at (-.5,-.8) {$1$};
        \draw[->] (0) -- (1);
        \node[untested] (2) at (-.5,-1.6) {$2$};
        \draw[->] (1) -- (2);
        \node[tested] (3) at (-.5,-2.4) {$3$};
        \draw[->] (2) -- (3);
        \draw[<-] (3) -- +(-.5,.4);
        \draw[->] (3) -- +(0,-.5);
        
        \node[tested] (4) at (.5,-.8) {$4$};
        \draw[->] (0) -- (4);
        \node[tested] (5) at (.5,-1.6) {$5$};
        \draw[->] (4) -- (5);
        \node[untested] (6) at (.2, -2.4) {$6$};
        \draw[->] (5) -- (6);
        \draw[->] (6) -- +(0, -.5);
        \node[tested] (7) at (.8, -2.4) {$7$};
        \draw[->] (5) -- (7);
        \draw[->] (7) -- +(0, -.5);
        
        \end{tikzpicture}
        \caption{Preemptive termination example}
        \label{fig:preemptive-termination}
    \end{wrapfigure}

    We now discuss the $\texttt{grow}$ function outlined in Algorithm \ref{tbl:path-tree}, which extends a given path-tree $\pt$ to depth $n$.
    The $\textit{covered}$ function, that is used on lines 5 and 10, returns 1 if the state $q$ has not been visited yet, and otherwise returns $0$.
    If the target depth $n$ and path-tree depth $d$ are equal, then no further calculations will be necessary and we immediately terminate (line 1 and 2).
    Otherwise, we first check whether $next$ is empty, i.e.\ no calculations belonging to this path-tree have been performed past the state $q$.
    If $\textit{next}$ is empty, a new path tree of depth $1$ is inserted for each transition originating from the state $q$ (lines 3 through 5).
    The $\texttt{grow}$ function is then called recursively on each path-tree in $\textit{next}$ using a target depth of $n-1$, and the variables $d$, $v$ and $v_{\textit{max}}$ are updated accordingly (lines 6 and 8 through 11).

    The \textbf{if} statement on line 7 corresponds to our second optimization which consists of pre-emptively terminating the calculation of a given path when we can determine that it will no longer be able to beat or the value of the current optimal path candidate.
    For example, take Figure \ref{fig:preemptive-termination} in which a part of a partially explored IOLTS is shown.
    For readability, the transition labels have been left out and previously visited states have been colored gray.
    The MBT algorithm is currently in state $0$ and wants to find an optimal path of length $3$.
    It first calculates a possible path of length $3$ along state $1$ and finds a possible path containing $2$ unvisited states.
    It then calculates a path along state $4$ but after seeing that the first two states, i.e.\ $4$ and $5$, have already been visited, the calculation is terminated since we can determine that the path along state $4$ can not contain $2$ or more unvisited states.
    
    \begin{algorithm}
        \caption{The greedy test case selection strategy.}
        \label{tbl:Strategy}
        \KwData{The IOLTS $L$, the set of currently maintained path-trees \textit{paths}, the set of current states $\textit{qs}$,
        the target depth of $n$, and a set of possible actions \textit{options}.}
        
        $\texttt{pick\_input}(L, \textit{paths}, \textit{qs}, n, \textit{options}) :=$ \;
        \tab{}{0} $\textit{paths} \gets \{ \textit{pt} \in \textit{paths} \; | \; \FuncSty{a}(\textit{pt}) \in \textit{options} \}$ \;
        \tab{}{0} \textbf{if} $\textit{paths} = \emptyset$ : \;
        \tab{}{1} \textbf{for} $q\in \textit{qs}$ \textbf{do:} \;
        \tab{}{2} \textbf{for} $a \in \textit{options}, q' \in Q$ \textbf{s.t.} $q \overset{a}{\to} q'$ \textbf{do:} \;
        \tab{}{3} \textbf{insert} $\langle a, q', \emptyset, 1, \textit{covered}(q'), 0 \rangle$ \textbf{in} \textit{paths}\;
        \tab{}{0} $v_{\textit{best}} \gets 0$ \;
        \tab{}{0} $\textit{pref} \gets \emptyset$ \;
        \tab{}{0} \textbf{for} $\textit{pt} \in \textit{paths} \textbf{ do:}$ \;
        \tab{}{1} $\texttt{grow}(L, \textit{pt}, n)$ \;
        \tab{}{1} \textbf{if} $\FuncSty{v}(\textit{pt}) > v_{\textit{best}}$ : \;
        \tab{}{2} $v_{\textit{best}} \gets \textit{path}.v$ \;
        \tab{}{2} $\textit{pref} \gets \{ \FuncSty{a}(\textit{pt}) \}$ \;
        \tab{}{1} \textbf{else if} $\FuncSty{v}(\textit{pt}) = v_{\textit{best}}$ : \;
        \tab{}{2} \textbf{insert} $\FuncSty{a}(\textit{pt})$ \textbf{in} \textit{pref} \;
        \tab{}{1} \textbf{return} \textit{pref} \;
    \end{algorithm}

    To use the greedy test-case selection strategy we make use of a variable \textit{paths} to store our current set of path trees originating from $qs$ and replace the statement on line $5$ of the MBT algorithm shown in Algorithm \ref{tbl:MBT_algorithm} with the following statement:
    \[\textbf{select } a_{\textit{in}} \textbf{ from } \texttt{pick\_input}(L, \textit{paths}, \textit{qs}, n, \nexta \cap A_I) \text{.}\]
    The \texttt{pick\_input} function is given in Algorithm \ref{tbl:Strategy} and takes care of growing all current path-trees to the target depth $n$, and reducing the set of possible input actions $\textit{options}$ to the ones belonging to a local-optimal path of length $n$.
    Whenever some action $a$ is performed by the implementation, i.e.\ an action $a$ is sent to or received from the implementation, the set of paths \textit{paths} is updated to the union of $next$ of all path trees whose action equals $a$.
    This allows us to reuse the already performed calculations.

%% file: sec6-benchmarks/sec6-main.tex
\section{Benchmarking results} \label{sec:results}
    We split up our benchmarks into two sets.
    In Section \ref{ssec:error_detection_rate}, we test whether our greedy strategy shows improvements over the random test case selection strategy similar to those of the global optimal solution provided in \cite{PetraTretmans2019}.
    This is done by benchmarking our strategy to the same set of benchmarks used for the global optimal solution.
    In Section \ref{ssec:scalability}, we benchmark the scalability of our solution by using both the greedy and the random strategy on a variety of large and complex generated and real-world models.

    For both sets of benchmarks, we are interested in the number of transitions each traversal strategy needs to reach a certain goal.
    For the first set, this goal is detecting non-conformance and for the second set, this goal is a specific percentage of state coverage.
    We measure the number of transitions instead of the amount of time required, since the throughput, i.e.\ the number of transitions per second, is largely dependent on the quiescence time, which can vary greatly per system.

    The systems that we look at in both sections are deterministic and internal-choice, i.e.\ in each state, it is possible to query inputs to the implementation or receive outputs from the implementation.

\subsection{Fault detection speed} \label{ssec:error_detection_rate}
    The benchmarks used in \cite{Tretmans2019} come from the automata wiki \cite{automata_wiki}, which is an online repository containing various formalized models that can be used for benchmarking.
    One set of these models pertains to the Bounded Retransmission Protocol \cite{BoundedRetransmissionProtocol} by Philips.
    For this communication protocol, the Wiki provides both a formalization of its intended behavior, as well as $6$ mutants.
    This protocol is a variation of the alternating-bit protocol and provides ordered and partially reliable communication for  a sequence split into three messages over a possibly unreliable communication channel.
    As such, a sequence of messages will always arrive in a clearly marked sequencing, however, the transmission of the entire sequence is terminated if the transmission of a given message has failed a fixed number of times.
    Each of the six mutants represents incorrect implementations of the communication protocol.
    We refer the reader to the wiki \cite{automata_wiki} for exact explanations of the differences between the mutants and the correct implementation. 

    As was done in \cite{PetraTretmans2019}, we measure the average amount of transitions required to find a non-conformance caused by each mutation using MBT for both the random strategy and our greedy strategy.
    For each strategy, the average is calculated over a total of $100$ runs per mutant.
    The averages of these runs, as well as the results by van den Bos and Tretmans \cite{PetraTretmans2019}, are shown in Table \ref{tbl:benchmarks_brp}.
    For reference, the last two columns contain the original benchmarks, where switch coverage is their global optimal solution and TorXakis is the symbolic random test case selection strategy.
    Since our tool does not allow for infinite data types, the variables representing the three messages have been replaced with constants.
    As such, we could not properly capture mutant six, since it makes assertions about the three messages that are used, and we omit this mutant from our benchmarks.
    We calculate the arithmetic mean of the averages of all 5 experiments for each test case selection strategy. 

    We observe a significant difference in the average detection speed when testing mutant 2 or 6.
    In mutant 2, the number of permitted failed transmissions after which communication is terminated is reduced from 5 to 4
    In mutant 5, an incorrect final response message is sent if only the first 2 messages have been transmitted correctly.
    Both mutations require a very specific sequence of inputs to become observable and, as the probability of such a specific sequence being picked at random is very low, the random test case selection strategies require significantly longer to observe these mutants.
    
    We observe that the greedy strategy performs $\approx 8.2$ times better than the random (non-symbolic) strategy.
    With the exclusion of Mutant 6, the global optimal solution actually performs $\approx 20.6$ times better than the random (symbolic) strategy instead of the $\approx 7.7$ times improvement that occurs when Mutant 6 is included.
    The original authors note that the decreased performance of the global optimal strategy applied to Mutant 6 is caused due the lack of randomness in deciding which values to use for the message variables.
    Therefore, we believe that the slight randomness of the greedy strategy would lead to similar performance results as those of random test case selection strategies and the performance improvement should stay the same as it is now if Mutant 6 were added to the set of benchmarks.
    
    \begin{table}[]
        \centering
        \begin{tabular}{|r|c|c||c|c|} \hline
         & \multicolumn{2}{c||}{Explicit data (2.9k states)} & \multicolumn{2}{c|}{Symbolic data (6 states)} \\ \hline
         & \begin{tabular}{c}
              greedy \\
              w. $N=5$ 
         \end{tabular} & random & A Priori & TorXakis\\ \hline \hline
        Mutant 1 & 20 & 22  & 44 & 12 \\
        Mutant 2 & 95 & 412 & 16 & 234 \\
        Mutant 3 & 16 & 21 & 8 & 12 \\
        Mutant 4 & 27 & 29 & 6 & 18 \\
        Mutant 5 & 180 & 2280 & 18 & 1620 \\ 
        Mutant 6 & - & - & 164 & 76 \\ \hline
        mean & 67.6 & 552.8 & 42.7 & 328.7\\ \hline
        \end{tabular}
        \caption{Average number of transitions required to different detect mutations of the Bounded Transmission Protocol though MBT using greedy test case selection, a priori generated path \cite{PetraTretmans2019}, and random test case selection}
        \label{tbl:benchmarks_brp}
    \end{table}
    
\subsection{Applicability on large models} \label{ssec:scalability}
    To investigate the effectiveness of our strategy when used on large systems, we measure the average amount of transitions required by both our greedy strategy and the random strategy, to cover a given percentage of states on such large models.
    These measurements are performed on both real-world models as well as generated models.
    We opt to also generate representative models since acquiring a large variety of large complex industrial models is rather difficult.
    The generated models are based on previous research by Groote et al.\ \cite{GROOTE201651}, in which random LTSs are generated and are shown to be representative of real-world models.
    This is done by comparing fault-detection results to real-life models and statistical analysis.
    
    The models generated in \cite{GROOTE201651}, are generated as follows:
    An LTS with $N$ states is generated, and each state is given $\lambda$ outgoing transitions with the target states being uniformly distributed.
    This process is repeated $p$ times, after which all $p$ random LTSs are parallel composed.
    The resulting model is then used as a representative LTS.
    
    The models which we generate, are meant to be representative of systems under run-to-completion semantics which dictates that once a system has begun processing some input, it will not start processing another input until the first input has been fully processed, even if both inputs could be processed simultaneously.
    These semantics are very useful as they significantly reduce the complexity of a model, both in terms of state-space as well as in terms of cognitive complexity.
    
    Since we can see the processing of a given input as a single monolithic action, we end up with similar-looking models to the aforementioned generated LTSs.
    We thus decide to generate our models as follows:
    We generate $p$ LTSs as was done before, each transition is labeled with some label representing an arbitrary input action followed by an arbitrary sequence of $r$ output action that forms the processing of the input action.
    Once all $p$ processes have been parallelly composed, the singular transitions are replaced with a sequence of transitions, labeled with the corresponding input action and the sequence of $r$ output actions.
    
    We require the resulting IOLTS to be deterministic and to be fully connected, i.e.\ there is always some path from each state to any other state.
    Using deterministic IOLTSs simplifies the problem situation, reduces variation during testing, and allows us to focus solely on the efficient traversal of the given model.
    The fully-connected requirement is imposed such that it is always possible to eventually test all states.
    Intuitively, a system that would not meet this requirement would favor an intelligent strategy that can avoid taking wrong transitions, i.e.\ a transition that causes yet untested states to become unreachable, in favor of the at-random approach.
    
    We implemented an LTS simulator, that mimics the behavior of the model, and use this as the implementation for all of our benchmarks since we are only interested in the exploration of our LTS.
    This eliminates a significant portion of the time spent on running the implementation and communication between this and the tool.
    
\begin{figure}
    \centering
    \fontsize{8}{10}\selectfont
    \includegraphics[width=0.8\textwidth]{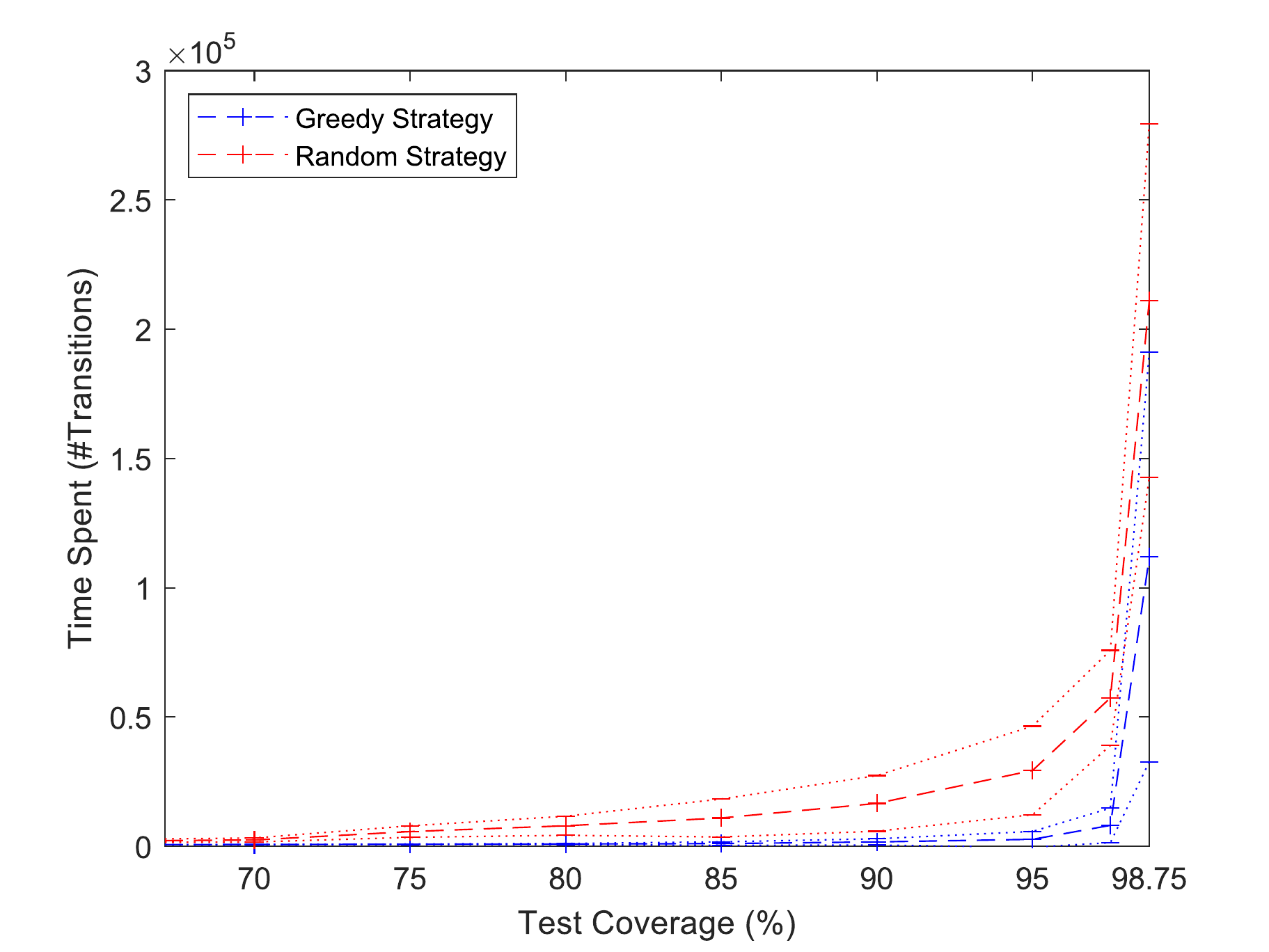}
    \caption{Number of transitions required to achieve given percentages of state coverage of the model at Philips by both the greedy and random strategies}
    \label{fig:mbt_on_philips}
    \end{figure}

    For our first set of benchmarks, we apply MBT on the model of an industrial component from Philips.
    This component is part of the larger set of software used on the Philips Azurion, a large and complex x-ray machine that allows for live imaging during critical heart-surgery operations.
    Figure \ref{fig:mbt_on_philips} shows the results of these benchmarks.
    The benchmarks consist of 10 runs per strategy.
    The average of these runs is indicated by the dashed line ($\relbar\relax\relbar$).
    The dotted lines ($\cdots$) show the average plus/minus the standard deviation of these runs.
    We see that the greedy strategy requires $\approx 6$ times fewer steps than the random strategy to cover $90\%$ of the states of the model at Philips.
    We note that the model contains transitions that, once traversed, cause certain states to no longer be reachable.
    As such, the random strategy could not consistently test more than $\approx 90\%$ of the states.
    
    For our second set of benchmarks, we use a subsystem of a large Dezyne specification of a sorting robot. 
    This model is part of the set of example models that come with the Dezyne tool, and we believe that, while being an example model, it is representative of machinery that can be encountered in the real world.
    The model is composed of a large set of modeled subcomponents and, as a result of this, consists of approximately $220\,000$ states.
    Since exploring such a large model would take a considerable amount of time, we use statespace reduction techniques to reduce its size whilst still maintaining all of the behavior.
    To do so, we make all quiescence labels explicit and then reduce our LTS modulo branching-bisimulation equivalence.
    This reduces the model from $220\,000$ states to only $617$ states.
    
\begin{figure}
    \centering
    \fontsize{8}{10}\selectfont
    \includegraphics[width=0.8\textwidth]{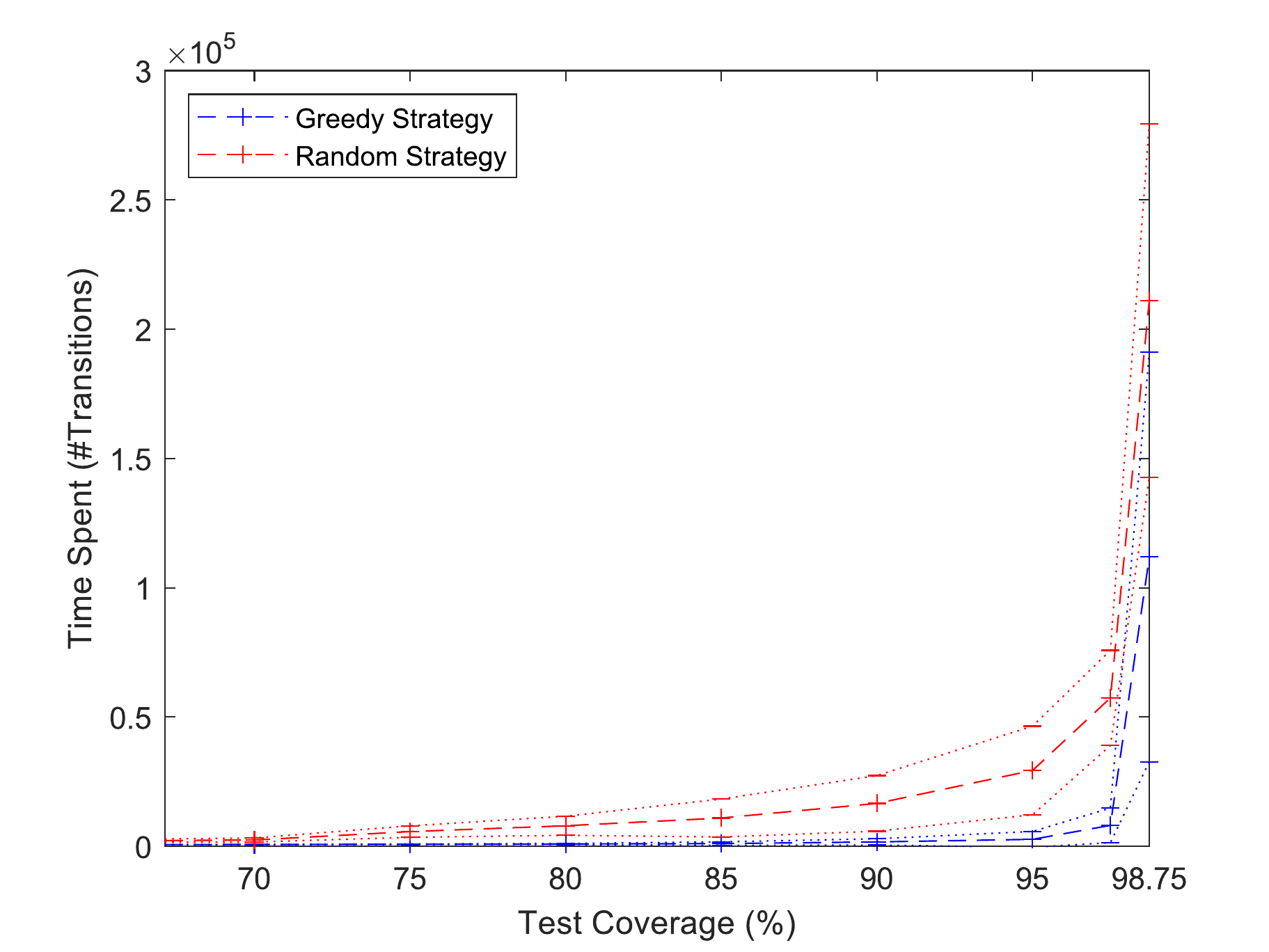}
    \caption{Number of transitions required to achieve given percentages of state coverage of the sorting machine model by both the greedy and random strategies}
    \label{fig:mbt_on_robot}
    \end{figure}

    Figure \ref{fig:mbt_on_robot} shows the results of the benchmarks performed on the Dezyne sorting machine and shows the greedy strategy requiring $\approx5$ times fewer transitions than the random strategy to cover $95\%$ of the sorting machine model.
    The benchmarks consist of $10$ runs per strategy.
    The average of these runs is indicated by the dashed line ($\relbar\relax\relbar$).
    The dotted lines (...) show the average plus/minus the standard deviation of these runs.
    
    For our final set of benchmarks, we apply both strategies to several randomly generated statespaces.
    All random statespaces are generated using the aforementioned technique.
    We used three different sets of generation parameters.
    Each set of parameters was used to generate $3$ random statespaces.
    Each randomly generated statespace was then used to perform 10 separate runs per strategy.
    The set of benchmarks thus consists of 30 runs per strategy per set of generation parameters.
    The statespaces used for the top-left benchmark are generated using the parameters $N=10, \lambda=6$, $r=1$, and $p=2$.
    The statespaces used for the top-right benchmark are generated using the parameters $N=10, \lambda=6$, $r=1$, and $p=3$.
    The statespaces used for the bottom benchmark are generated using the parameters $N=800, \lambda=6, r=1$, and $p=1$.
        
    \begin{figure}
        \centering
        \fontsize{8}{10}\selectfont
        \includegraphics[width=0.8\textwidth]{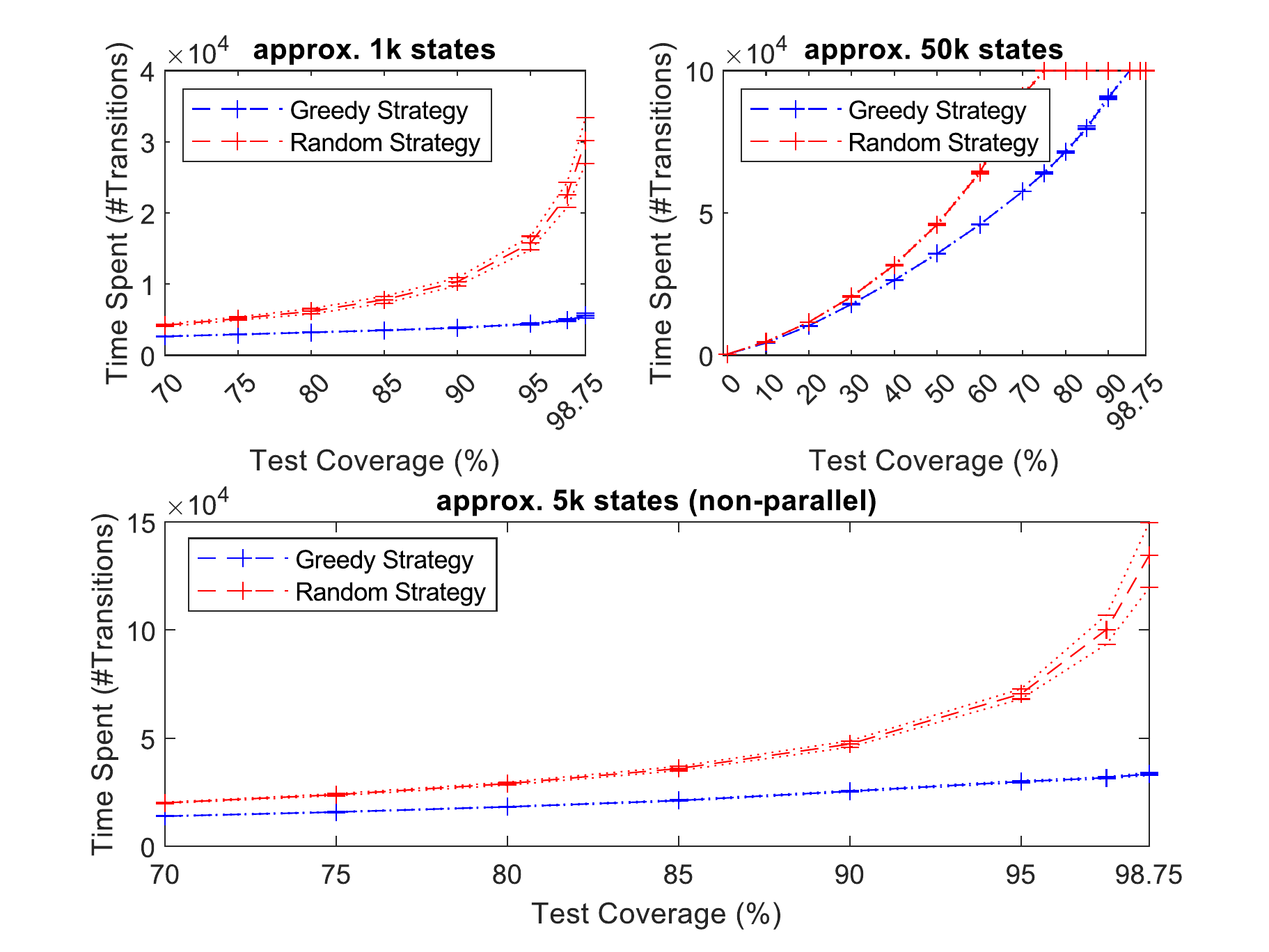}
        \caption{Number of transitions required to achieve given percentages of state coverage of several generated statespaces}
        \label{fig:mbt_on_gen}
    \end{figure}

    Figure \ref{fig:mbt_on_gen} shows the results of the benchmarks performed on the generated statespaces.
    The greedy strategy, on average, required $\approx 5.5 \times$ fewer transitions than the random strategy to reach $98.75\%$ of all states, significantly fewer transitions on the much larger statespaces that were used in the top-right benchmarks, and $\approx 4.2 \times$ fewer transitions on the non-parallel statespaces used for the bottom benchmarks.
    All of the benchmarks showcase similar improvements over the old random strategy.
    

%% file: sec7-conclusion/sec7-main.tex
\section{Conclusion and future work} \label{sec:fin}
We conclude that our greedy solution is a significant improvement over the old at-random exploration strategy when we want to thoroughly test complex systems using dynamic MBT.
Our strategy has shown to have a significant speedup over the random strategy when applied to both the industrial model and the sorting machine, as well as the large set of generated statespaces.

However, how often we can query the tested implementation within a given timeframe still remains to be a big bottleneck for MBT. And we believe this to be an important area for future research if we want to be able to efficiently use MBT in software development.
More future work lies in studying non-fully connected models, in which a wrong decision could cause a significant portion of states to no longer be visitable.
One might propose to solve this, by allowing the test suite to reset the system.
Resetting real-world systems is however time-consuming, and thus knowing when to trigger a reset could be crucial in speeding up testing. 
Another interesting aspect of such systems is finding and comparing strategies that aim to maximize the possible state coverage, by avoiding such wrong decisions.

%% file: main.bbl
\begin{thebibliography}{10}
\providecommand{\url}[1]{\texttt{#1}}
\providecommand{\urlprefix}{URL }
\providecommand{\doi}[1]{https://doi.org/#1}

\bibitem{AICHERNIG2015383}
Aichernig, B.K., Jöbstl, E., Tiran, S.: Model-based mutation testing via
  symbolic refinement checking. Science of Computer Programming  \textbf{97},
  383--404 (2015). \doi{https://doi.org/10.1016/j.scico.2014.05.004}, special
  Issue: Selected Papers from the 12th International Conference on Quality
  Software (QSIC 2012)

\bibitem{SearchOptimization}
Araujo, H., Carvalho, G., Mousavi, M.R., Sampaio, A.: Multi-objective search
  for effective testing of cyber-physical systems. In: {\"O}lveczky, P.C.,
  Sala{\"u}n, G. (eds.) Software Engineering and Formal Methods. pp. 183--202.
  Springer International Publishing, Cham (2019)

\bibitem{beck2003test}
Beck, K.: Test-driven development: by example. Addison-Wesley Professional
  (2003)

\bibitem{BJN07}
Best, B., Jurjens, J., Nuseibeh, B.: Model-based security engineering of
  distributed information systems using umlsec. In: 29th International
  Conference on Software Engineering (ICSE'07). pp. 581--590 (2007).
  \doi{10.1109/ICSE.2007.55}

\bibitem{TDD}
Bhat, T., Nagappan, N.: Evaluating the efficacy of test-driven development:
  Industrial case studies. In: Proceedings of the 2006 ACM/IEEE International
  Symposium on Empirical Software Engineering. p. 356–363. ISESE '06,
  Association for Computing Machinery, New York, NY, USA (2006).
  \doi{10.1145/1159733.1159787}

\bibitem{PetraTretmans2019}
van~den Bos, P., Tretmans, J.: Coverage-based testing with symbolic transition
  systems. In: Beyer, D., Keller, C. (eds.) Tests and Proofs. pp. 64--82.
  Springer International Publishing, Cham (2019)

\bibitem{Cartaxo11}
Cartaxo, E., Machado, P., de~Oliveira~Neto, F.: On the use of a similarity
  function for test case selection in the context of model-based testing.
  Softw. Test., Verif. Reliab.  \textbf{21},  75--100 (06 2011).
  \doi{10.1002/stvr.413}

\bibitem{BoundedRetransmissionProtocol}
Groote, J.F., van~de Pol, J.: A bounded retransmission protocol for large data
  packets. In: Algebraic Methodology and Software Technology. pp. 536--550.
  Springer Berlin Heidelberg (1996)

\bibitem{GROOTE201651}
Groote, J.F., {van der Hofstad}, R., Raffelsieper, M.: On the random structure
  of behavioural transition systems. Science of Computer Programming
  \textbf{128},  51--67 (2016).
  \doi{https://doi.org/10.1016/j.scico.2016.02.006},
  \url{https://www.sciencedirect.com/science/article/pii/S0167642316000599},
  special issue on Automated Verification of Critical Systems (AVoCS’14)

\bibitem{Henzinger2000}
Henzinger, T.A.: The Theory of Hybrid Automata, pp. 265--292. Springer Berlin
  Heidelberg, Berlin, Heidelberg (2000). \doi{10.1007/978-3-642-59615}

\bibitem{funcsearch}
Lefticaru, R., Ipate, F.: Functional search-based testing from state machines.
  In: 2008 1st International Conference on Software Testing, Verification, and
  Validation. pp. 525--528 (2008). \doi{10.1109/ICST.2008.32}

\bibitem{MiL-spacereduction}
Matinnejad, R., Nejati, S., Briand, L., Brcukmann, T.: Mil testing of highly
  configurable continuous controllers: Scalable search using surrogate models.
  In: Proceedings of the 29th ACM/IEEE International Conference on Automated
  Software Engineering. p. 163–174. ASE '14, Association for Computing
  Machinery, New York, NY, USA (2014). \doi{10.1145/2642937.2642978}

\bibitem{automata_wiki}
Neider, D., Smetsers, R., Vaandrager, F., Kuppens, H.: Benchmarks for automata
  learning and conformance testing (2019). \doi{10.1007/978-3-030-22348-9-23}

\bibitem{IOCO}
Tretmans, J.: Model Based Testing with Labelled Transition Systems, pp. 1--38.
  Springer Berlin Heidelberg, Berlin, Heidelberg (2008).
  \doi{10.1007/978-3-540-78917-8}

\bibitem{Tretmans2019}
Tretmans, J., van~de Laar, P.: Model-based testing with torxakis. Central
  European Conference on Information and Intelligent Systems pp. 247--258
  (2019),
  \url{https://www.proquest.com/conference-papers-proceedings/model-based-testing-with-torxakis/docview/2366658121/se-2}

\end{thebibliography}
